\documentclass[a4paper,fleqn,usenatbib]{mnras}

\usepackage{newtxtext,newtxmath}

\usepackage[T1]{fontenc}
\usepackage{ae,aecompl}

\usepackage{graphicx}	
\usepackage{amsmath}	
\usepackage{amssymb}	
\usepackage{mathrsfs}	
\usepackage{natbib}
\usepackage{color}

\newcommand{\sdssu}{\hbox{$u$}}
\newcommand{\sdssg}{\hbox{$g$}}
\newcommand{\sdssr}{\hbox{$r$}}
\newcommand{\sdssi}{\hbox{$i$}}
\newcommand{\sdssz}{\hbox{$z$}}

\newcommand{\mstar}{\hbox{$M_{\star}$}}

\newcommand{\msol}{\hbox{$M_\odot$}}

\newcommand{\kms}{\hbox{km~s$^{-1}$}}

\newcommand{\permpc}{\hbox{Mpc$^{-1}$}}

\newcommand{\perpixel}{\hbox{pixel$^{-1}$}}

\newcommand{\eg}{e.g.}
\newcommand{\ie}{i.e.}
\newcommand{\citeeg}[1]{(\eg, \citealt{#1})}

\title[Galaxy metallicity with CNNs]{Using convolutional neural networks to predict galaxy metallicity from three-color images}

\author[Wu and Boada]
{\parbox{\textwidth}{John~F.~Wu$^{1}$\thanks{E-mail: jfwu@physics.rutgers.edu} and
Steven~Boada$^{1}$}\vspace{0.4cm}\
\\
\parbox{\textwidth}{$^{1}$Department of Physics and Astronomy, Rutgers, The State University of New Jersey, 136 Frelinghuysen Road, Piscataway, NJ 08854-8019, USA\\}}

\date{Accepted XXX. Received YYY; in original form ZZZ}

\pubyear{2019}

\begin{document}
\label{firstpage}
\pagerange{\pageref{firstpage}--\pageref{lastpage}}
\maketitle

\begin{abstract}
We train a deep residual convolutional neural network (CNN) to predict the gas-phase metallicity ($Z$) of galaxies derived from spectroscopic information ($Z \equiv 12 + \log(\rm O/H)$) using only three-band \sdssg\sdssr\sdssi\ images from the Sloan Digital Sky Survey. When trained and tested on $128 \times 128$-pixel images, the root mean squared error (RMSE) of $Z_{\rm pred} - Z_{\rm true}$ is only 0.085~dex, vastly outperforming a trained random forest algorithm on the same data set (RMSE $=0.130$~dex). The amount of scatter in $Z_{\rm pred} - Z_{\rm true}$ decreases with increasing image resolution in an intuitive manner. We are able to use CNN-predicted $Z_{\rm pred}$ and independently measured stellar masses to recover a mass-metallicity relation with $0.10$~dex scatter. Because our predicted MZR shows no more scatter than the empirical MZR, the difference between $Z_{\rm pred}$ and $Z_{\rm true}$ can not be due to purely random error. This suggests that the CNN has learned a representation of the gas-phase metallicity, from the optical imaging, beyond what is accessible with oxygen spectral lines.
\end{abstract}


\section{Introduction}\label{sec:introduction}
Large-area sky surveys, both on-going and planned, are revolutionizing our understanding of galaxy evolution. The Dark Energy Survey (DES; \citealt{DES2005}) and upcoming Large Synoptic Survey Telescope (LSST; \citealt{LSST2012}) will scan vast swaths of the sky and create samples of galaxies of unprecedented size. Spectroscopic follow-up of these samples will be instrumental in order to understand their properties. Previously, the Sloan Digital Sky Survey \citep[SDSS;][]{York2000} and its spectroscopic campaign enabled characterization of the mass-metallicity relation (hereafter MZR; \citealt{Tremonti2004}) and the fundamental metallicity relation, (hereafter FMR; \eg, \citealt{Mannucci2010}). As future surveys are accompanied by larger data sets, individual spectroscopic follow-up observations will become increasingly impractical.

Fortunately, the large imaging data sets to be produced are ripe for application of machine learing (ML) methods. ML is already showing promise in studies of galaxy morphology \citeeg{Dieleman2015, Huertas-Company2015, Beck2018, Dai2018, Hocking2018}, gravitational lensing \citeeg{Hezaveh2017, Lanusse2017, Petrillo2017, Petrillo2018}, galaxy clusters \citeeg{Ntampaka2015, Ntampaka2016}, star-galaxy separation \citeeg{Kim2017}, creating mock galaxy catalogs \citeeg{Xu2013}, asteroid identification \citeeg{Smirnov2017}, and photometric redshift estimation \citeeg{Hoyle2016, DIsanto2018, Pasquet2019}, among many others. ML methods utilizing neural networks have grown to prominence in recent years. While neural networks are a relatively old technique \citeeg{LeCun1989}, their recent increase in popularity is driven by the widespread availability of affordable graphics processing units (GPUs) that can be used to do general purpose, highly parallel computing. Also, unlike more ``traditional'' ML methods, neural networks excel at image classification and regression problems.

Inferring spectroscopic properties from the imaging taken as part of a large-area photometric survey is, at a basic level, an image regression problem. These problems are most readily solved by use of convolutions in multiple layers of the network (see, \eg, \citealt{Krizhevsky2012}). Convolutional neural networks (CNNs, or convnets) efficiently learn spatial relations in images whose features are about the same sizes as the convolution filters (or kernels) that are to be learned through training. CNNs are considered \textit{deep} when the number of convolutional layers is large. Visualizing their filters reveals that increased depth permits the network to learn more and more abstract features (\eg, from Gabor filters, to geometric shapes, to faces; \citealt{Zeiler2014}).

In this work, we propose to use supervised ML by training CNNs to analyze pseudo-three color images and predict the gas-phase metallicity. We use predicted metallicities to recover the empirical \cite{Tremonti2004} MZR. This paper is organized as follows: In Section~\ref{sec:data}, we describe the acquisition and cleaning of the SDSS data sample. In Section~\ref{sec:training}, we discuss selection of the network's hyperparameters and outline training the of network. We present the main results in Section~\ref{sec:results}. In Section~\ref{sec:interpretation}, we interpret the CNN's performance and discuss our findings in the context of current literature. In Section~\ref{sec:MZR}, we characterize the MZR using the metallicity predicted by our CNN. We summarize our key results in Section~\ref{sec:summary}.

Unless otherwise noted, throughout this paper, we use a concordance cosmological model ($\Omega_\Lambda = 0.7$, $\Omega_m = 0.3$, and $H_0= 70$ \kms{} \permpc), assume a Kroupa initial mass function \citep{Kroupa2001}, and use AB magnitudes \citep{Oke1974}.

\section{Data} \label{sec:data}
To create a large training sample, we select galaxies from the Sloan Digital Sky Survey (SDSS; \citealt{York2000}) DR7 MPA/JHU spectroscopic catalog \citep{Kauffmann2003a, Brinchmann2004, Tremonti2004, Salim2007}. The catalog provides spectroscopically derived properties such as stellar mass (\mstar) and gas-phase metallicity ($Z$) estimates \citep{Tremonti2004}. We select objects with low reduced chi-squared of model fits (\texttt{rChi2} $< 2$), and median $Z$ estimates available (\texttt{oh\_p50}). We supplement the data from the spectroscopic catalog with photometry in each of the five SDSS photometric bands (\sdssu, \sdssg, \sdssr, \sdssi, \sdssz), along with associated errors from SDSS DR14 \citep{Abolfathi2017}.

We require that galaxies magnitudes are $10 < \sdssu \sdssg \sdssr \sdssi \sdssz < 25$ mag, in order to avoid saturated and low signal-to-noise detections. We enforce a color cut, $0 < \sdssu-\sdssr < 6$, in order to avoid extremely blue or extremely red objects, and require objects to have spectroscopic redshifts greater than $z=0.02$ with low errors ($z_{err} < 0.01$). The median redshift is $0.07$ and the highest-redshift object has $z = 0.38$. We also require that the \sdssr-band magnitude measured inside the Petrosian radius (\texttt{petroMag\_r}; \citealt{Petrosian1976}) be less than 18 mag, corresponding to the spectroscopic flux limit. With these conditions we construct an initial sample of 142,182 objects (there are four objects with duplicate SDSS DR14 identifiers).
We set aside 25,000 objects for later testing, and use the rest for training and validation.

We create RGB image cutouts of each galaxy with the SDSS cutout service\footnote{\url{http://skyserver.sdss.org/dr14/en/help/docs/api.aspx}}, which converts \sdssg\sdssr\sdssi\ bands to RGB channels according to the algorithm described in \cite{Lupton2004} (with modifications by the SDSS SkyServer team). Since images are not always available, we are left with 116,429 SDSS images with metallicity measurements, including 20,466/25,000 of the test subsample. We create $128\times128$-pixel JPG images with a pixel scale of $0\farcs296$, which corresponds to $38''\times 38''$ on the sky.  We do not further preprocess, clean, or filter the images before using them as inputs to our CNN.

\section{Methodology}\label{sec:training}
Before the CNN can be asked to make predictions, it must be trained to learn the relationships between the input data (the images described above) and the desired output (metallicity). The CNN makes predictions using the input images, and the error (or loss) is determined based on the differences between true and predicted values. The CNN then updates its parameters, or weights, in a way that minimizes the loss function. We use the root mean squared error loss function:
\begin{equation} \label{eq:rmse}
\mathrm{RMSE} \equiv \sqrt{\langle |y_{\rm true} - y_{\rm pred}|^2\rangle},
\end{equation}
where $y_{\rm true}$ is the ``true'' and $y_{\rm pred}$ is the predicted value, and $y$ represents the target quantity.

It is worth emphasizing that the $Z_{\rm true}$ is the metallicity estimated by model fits to strong emission lines in the SDSS spectra. \cite{Tremonti2004} determine a likelihood distribution of metallicities based on the model fits, and we define their 50th percentile metallicity estimates to be the \textit{true} metallicity ($Z_{\rm true}$) for the purpose of training our network. The typical systematic uncertainty in their metallicity model fits is about 0.03~dex.

We randomly split our training sample of $\sim 96,953$ images into 80\% (76,711) training and 20\% (19,192) validation data sets, respectively. The test data set of 20,466 images is isolated for now, and is not accessible to the CNN until all training is completed. Images and $Z_{\rm true}$ answers are given to the CNN in ``batches'' of 256 at a time, until the full training data set has been used for training. Each full round of training using all of the data is called an epoch, and we compute the loss using the validation data set at the end of each epoch. We use gradient descent for each batch to adjust weight parameters, and each weight's fractional contribution of loss is determined by the backpropagation algorithm \citep{LeCun1989}, during which finite partial derivatives are computed and propagated backwards through layers (i.e., using the chain rule for derivatives).

We use a 34-layer residual CNN architecture \citep{He2015} initialized to weights pre-trained on the ImageNet data set, which consists of 1.7 million images belonging to 1000 categories of objects found on Earth \citep[e.g., cats, horses, cars, or books;][]{ImageNet}. The CNN is trained for a total of 10 epochs. For more details about the CNN architecture, transfer learning, hyperparameter selection, data augmentation, and the training process, see the Appendix. In total, our training process requires 25-30 minutes on our GPU and uses under 2~GB of memory.

We evaluate predictions using the RMSE loss function, which approaches the standard deviation for Gaussian-distributed data. We also report the NMAD, or the normal median absolute deviation \citeeg{Ilbert2009, Dahlen2013, Molino2017}:
\begin{equation}
{\rm NMAD}(x) \approx 1.4826 \times {\rm median} \big (\big|x - {\rm median }(x) \big|\big ),
\end{equation}
where for a Gaussian-distributed $x$, the NMAD will also approximate the standard deviation, $\sigma$.
NMAD has the distinct advantage in that it is insensitive to outliers and can be useful for measuring scatter.
However, unlike the RMSE, which quantifies the typical scatter distributed about a center of zero, NMAD only describes the scatter around the (potentially non-zero) median.

\section{Results}\label{sec:results}

\subsection{Example predictions}
\begin{figure*}
	\includegraphics[width=0.9\textwidth]{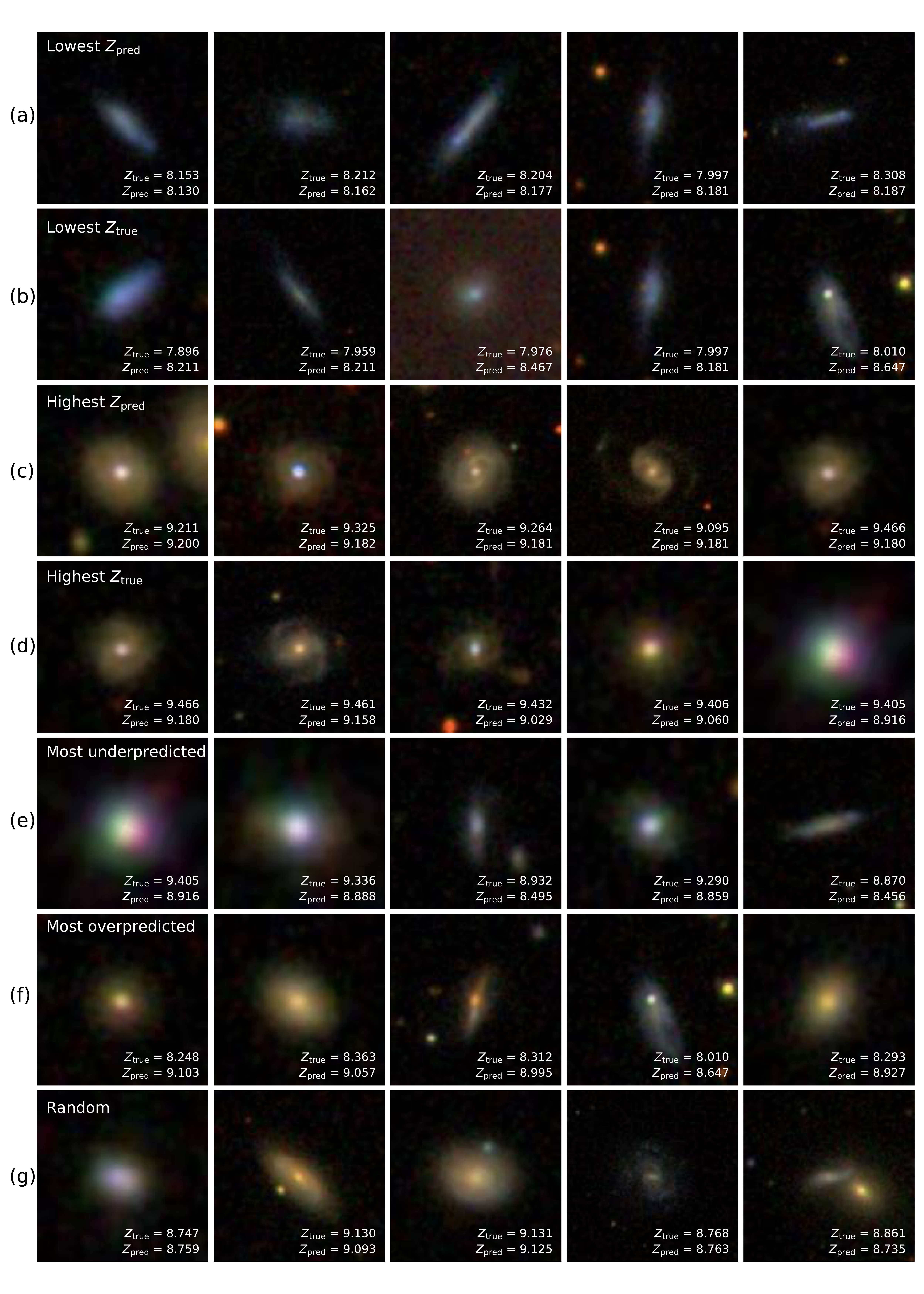}
	\caption{\label{fig:examples}
		SDSS imaging with predicted and true metallicities from the test data set. Five examples are shown from each of the following categories: (a) lowest predicted metallicity, (b) lowest true metallicity, (c) highest predicted metallicity, (d) highest true metallicity, (e) most under-predicted metallicity, (f) most over-predicted metallicity, and (g) a set of randomly selected galaxies.}
\end{figure*}

In Figure~\ref{fig:examples}, we show examples of $128 \times 128$ pixel \sdssg\sdssr\sdssi\ SDSS images that are evaluated by the CNN. Rows (a) and (b) depict the galaxies with lowest predicted and lowest true metallicities, respectively. The CNN associates blue, edge-on disk galaxies with low metallicities, and is generally accurate in its predictions. In rows (c) and (d), we show the galaxies with highest predicted and highest true metallicities, respectively. Here we find that red galaxies containing prominent nuclei are predicted to be high in metallicity, and that their predictions generally match $Z_{\rm true}$.

Galaxies predicted by our CNN to have high metallicities ($Z_{\rm pred} > 9.0$) tend to be characterized by high $Z_{\rm true}$, and the equivalent is true for low-metallicity galaxies. Conversely, galaxies with the highest (lowest) \textit{true} metallicities in the sample are also predicted to have high (low) metallicities. Note that inclined galaxies tend to be lower in metallicity whereas face-on galaxies appear to be higher in metallicity. \cite{Tremonti2004} explain this correlation by suggesting that the SDSS fiber aperture captures more column of a projected edge-on disk, allowing the metal-poor, gas-rich, and less-extincted outer regions to more easily be detected and depress the integrated $Z_{\rm true}$.

We will now consider examples of the most incorrectly predicted galaxies. In rows (e) and (f), we show instances in which the CNN predicted too low metallicity and too high metallicity, respectively. The two galaxies with the most negative residuals $\Delta Z \equiv Z_{\rm pred} - Z_{\rm true}$ (\ie, most under-predicted metallicities) suffer from artifacts that cause unphysical color gradients, and/or are labeled as quasars on the basis of their SDSS spectra (for which we expect $Z_{\rm true}$ to be biased). It is not unsurprising that the CNN has made mistakes in some of these cases, since they go against astronomers' usual heuristics: blue, disk-dominated sources are generally thought of as lower in metallicity, and redder, more spheroidal objects tend to be higher in metallicity.

In the bottom row (g) of Figure~\ref{fig:examples}, we show five randomly selected galaxies. The random SDSS assortment consists of lenticular, spiral, and possibly even an interacting pair of galaxies. Residuals are low (below 0.15~dex), and we again find that the CNN predictions track with human visual intuition.

\subsection{Comparing predicted and true metallicities}
\begin{figure}
	\includegraphics[width=\columnwidth]{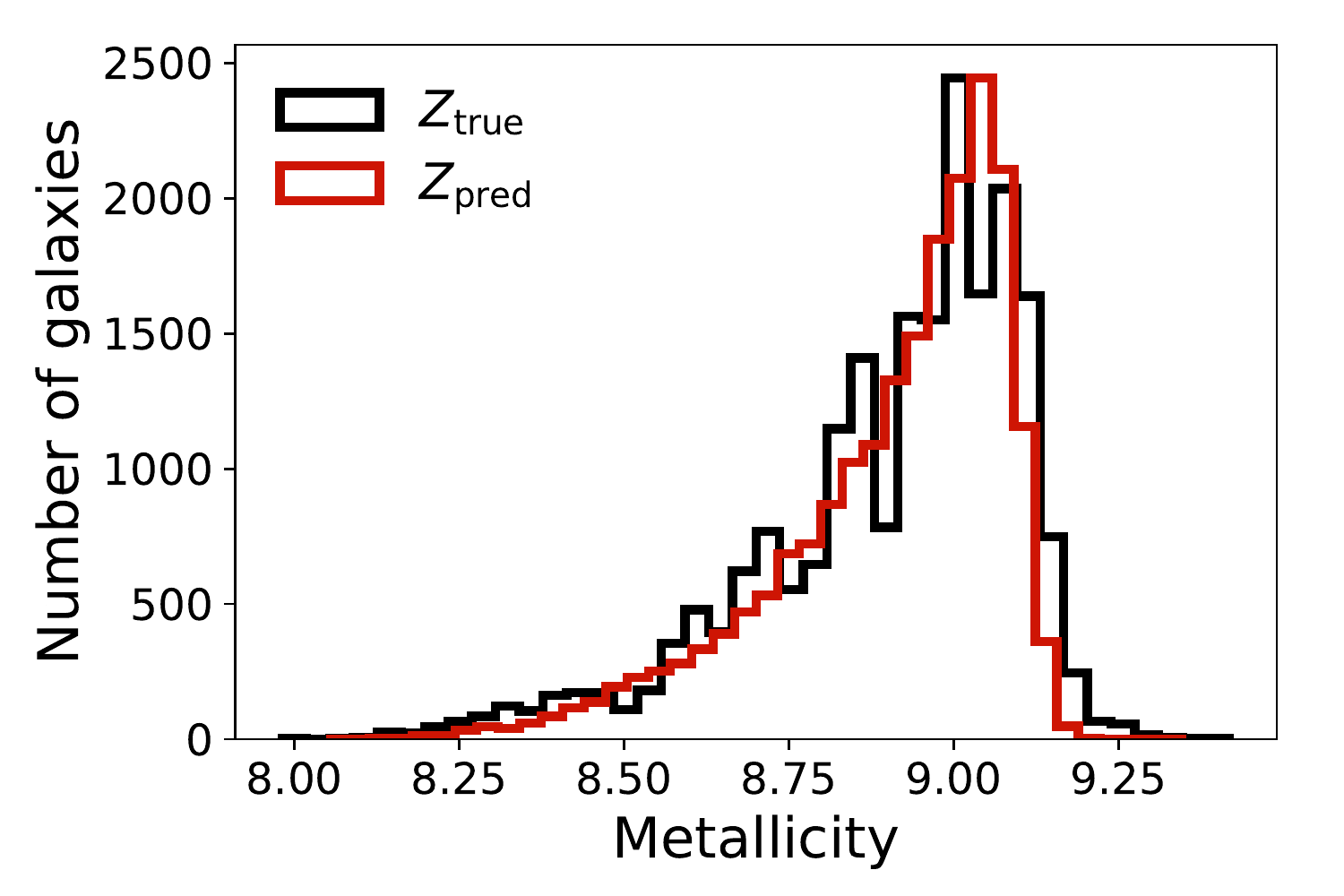}
	\caption{\label{fig:distributions}
		Distributions of the true (black) and predicted (red) galaxy metallicities. Note that the bin widths are different for the two distributions. See text for details.}
\end{figure}

In Figure~\ref{fig:distributions}, we show histograms of the true and predicted metallicities in black and red, respectively. The histogram bin sizes are chosen according to the \cite{Freedman1981} rule for each distribution. The discreet striping of the \cite{Tremonti2004} and \cite{Brinchmann2004} metallicity estimator appears in the $Z_{\rm true}$ distribution but does not appear in our CNN predictions. This striping should increase the scatter in our distribution of residuals.

The range of $Z_{\rm pred}$ is more limited than the range of $Z_{\rm true}$, which can also be seen from Figure~\ref{fig:examples} for extreme values of $Z_{\rm true}$. Too narrow a domain in $Z_{\rm pred}$ will lead to systematic errors, as the CNN will end up never predicting very high or very low metallicities. Although the two distributions are qualitatively consistent with each other at low metallicities (\eg, $Z < 8.5$), the fraction of galaxies with high $Z_{\rm true} > 9.1$ ($2573/20466 = 12.6\%$) is higher than the fraction with high $Z_{\rm pred} > 9.1$ ($1174/20466 = 5.7\%$).

We find that the mode of the binned predicted metallicity distribution is higher than that of $Z_{\rm true}$. This result may be a consequence of the CNN overcompensating for its systematic under-prediction of metallicity for galaxies with $Z_{\rm true} > 9.1$. However, its effect on the entire distribution is small, and may be remedied simply by increasing the relative fraction of very high-$Z_{\rm true}$ objects. We find overall good qualitative agreement between the $Z_{\rm pred}$ and $Z_{\rm true}$ distributions.

\subsection{Scatter in $Z_{\rm pred}$ and $Z_{\rm true}$}
\begin{figure}
	\includegraphics[width=\columnwidth]{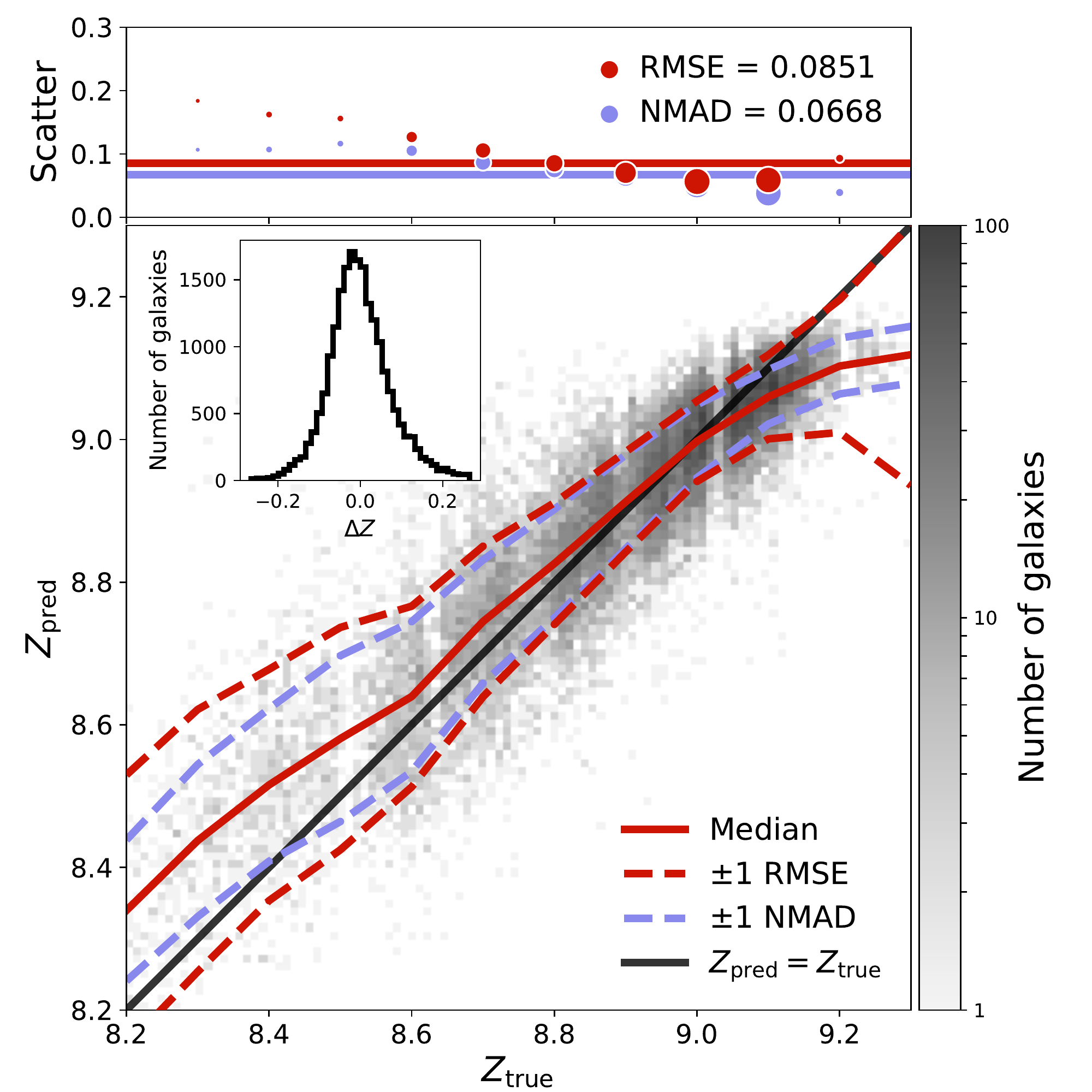}
	\caption{\label{fig:predicting-metallicity}
		Bivariate distribution of true galaxy metallicity ($Z_{\rm true}$) and CNN prediction ($Z_{\rm pred}$) is shown in the main panel. Overlaid are the median predicted metallicity (solid red line), RMSE scatter (dashed red lines), and NMAD scatter (dashed violet lines), in bins of $Z_{\rm true}$. The solid black line shows the one-to-one relation. The distribution of residuals ($Z_{\rm pred} - Z_{\rm true}$) is shown in the inset plot. In the upper panel, we again show the binned scatter, where the size of each marker is proportional to the number of galaxies in that bin. Each horizontal line corresponds to the average scatter over the entire test data set (and the global value indicated in the upper panel legend).}
\end{figure}

In Figure~\ref{fig:predicting-metallicity}, we compare the distributions of $Z_{\rm true}$ and $Z_{\rm pred}$ using a two-dimensional histogram (shown in grayscale in the main, larger panel). We also show the median predictions varying with binned $Z_{\rm true}$ (solid red line), in addition to the scatter in RMSE (dashed red) and NMAD (dashed violet), and also the one-to-one line (solid black). The running median agrees well with the one-to-one line, although at low metallicity we find that the CNN makes makes overpredictions.

A histogram of metallicity residuals is shown in the inset plot of the Figure~\ref{fig:predicting-metallicity} main panel. The $\Delta Z$ distribution is characterized by an approximately normal distribution with a heavy tail at large positive residuals; this heavy tail is likely due to the systematic over-prediction for low-$Z_{\rm true}$ galaxies.
There is also an overabundance of large negative $\Delta Z$ corresponding to under-predictions for high $Z_{\rm true}$, although this effect is smaller. We do not find significant correlations between $\Delta Z$ and galaxy observables including spectroscopic redshift, any combination of photometric color (including \sdssu\ and \sdssz\ bands), emission line signal-to-noise ratios, observed \sdssg\sdssr\sdssi-magnitudes, or axis ratios.

We now turn our attention to the upper panel of Figure~\ref{fig:predicting-metallicity}, which shows how the scatter varies with spectroscopically derived metallicity. The RMSE scatter and outlier-insensitive NMAD are both shown. Marker sizes are proportional in area to the number of samples in each $Z_{\rm true}$ bin, and the horizontal lines are located at the average loss (RMSE or NMAD) for the full test data set.

Predictions appear to be both accurate and low in scatter for galaxies with $Z_{\rm true} \approx 9.0$, which is representative of a typical metallicity in the SDSS sample. Where the predictions are systematically incorrect, we find that the RMSE increases dramatically. However, the same is not true for the NMAD; at $Z_{\rm true} < 8.5$, it asymptotes to $\sim 0.10$~dex, even though the running median is incorrect by approximately the same amount! This discrepancy is because the NMAD determines the scatter about the \textit{median} and not $\Delta Z = 0$, and thus, this metric becomes somewhat unreliable when the binned samples do not have a median value close to zero. Fortunately, the global median of $\Delta Z$ is $-0.006$~dex, or less than 10\% of the RMSE, and thus the global NMAD $= 0.067$~dex is representative of the outlier-insensitive scatter for the entire test data set.

This effect partly explains why the global NMAD ($0.067$~dex) is higher than the weighted average of the binned NMAD ($\sim 0.05$~dex). Also, each binned NMAD is computed using its local scatter, such that the outlier rejection criterion varies with $Z_{\rm true}$. To illustrate this effect with an example: $\Delta Z \approx 0.2$~dex would be treated as an $3\,\sigma$ outlier at $Z_{\rm true} = 9.0$, where the CNN is generally accurate, but the same residual would not be rejected as an outlier using NMAD for $Z_{\rm true} = 8.5$. Since the binned average NMAD depends on choice of bin size, we do not include those results in our analysis and only focus on the global NMAD. RMSE is a robust measure of both local and global scatter (although it becomes biased high by outliers).

\section{Interpreting the CNN} \label{sec:interpretation}

\subsection{Uncertainty in the scatter} \label{sec:scatter-uncertainties}

It is worth examining how reliable our estimate of RMSE$~=0.085$~dex is. Because we have a large data set, we can calculate uncertainties on the RMSE through multiple training/test realizations. 
One possible method is by dividing our full data set into cross-validation and nested cross-validation splits in order to see how the RMSE varies.

For the first method (5-fold cross-validation), we take the entire data set from Section~\ref{sec:data} and split it into five 80\%/20\% training/test subsets, each of which is optimized independently. We then compute the mean and standard deviation of the five test samples' $Z_{\rm pred}$, and find that the RMSE~$= 0.0836 \pm 0.0005$. Because there are more training examples here than in our original training set, the mean RMSE is lower than what we have previously found in Section~\ref{sec:results}.

For the second method (nested cross-validation), we split the full data set into five 80\%/20\% training/validation-test splits, and then further divide the training/validation data sets into 75\%/25\% cross-validation splits. We compute the mean and standard deviation of $Z_{\rm pred}$ for the ensemble of 5-fold test splits. When we select the model with the best cross-validation score, the RMSE is $0.0823 \pm 0.0009$. The unweighted average of all training/validation models is RMSE~$=0.0831 \pm 0.0011$.

There is an additional source of scatter due to noise in the SDSS images' pixels. This noise is not uniform across SDSS images, and the \cite{Lupton2004} intensity scaling makes estimating or re-sampling the noise distribution challenging. Therefore, we do not account for the contribution of image noise to our estimate of the uncertainties.



\subsection{Impact of artificially increasing scatter} \label{sec:additional-scatter}

In order to simulate additional uncertainty that may arise from noisier measurements of spectral lines, we add normally distributed scatter to the $Z_{\rm true}$ values. We train our CNN as before (in Section~\ref{sec:results}), except that we add random $\sigma = \{0.03, 0.05, 0.10, 0.20\}$~dex of scatter to the target $Z_{\rm true}$. The smallest value, $0.03~$dex, is the same as the systematic uncertainty in the \cite{Tremonti2004} measurements, and $0.20$~dex represents the standard deviation for the entire $Z_{\rm true}$ distribution. We compare CNN-predicted $Z_{\rm pred}$ with the original $Z_{\rm true}$ (i.e., the underlying values without artificial scatter introduced) using the RMSE metric as before (Equation~\ref{eq:rmse}). For all values of additional scatter, the CNN is able to estimate $Z_{\rm pred}$ to RMSE~$ = \{0.0851, 0.0851, 0.0869, 0.0882\}$~dex respectively. These results show that the CNN is robust to extra scatter added in an unbiased way. 

As a second test, we include random scatter drawn from a normal distribution centered at 0 and with standard deviation equal to the galaxy's redshift. This simple model can test how the CNN responds to redshift-dependent scatter in $Z_{\rm true}$. We note that this toy model disproportionately impacts higher-metallicity galaxies because our sample of lower-mass (and thus, lower-metallicity) galaxies is less complete at higher redshifts. After training on the original images with these modified $Z_{\rm true}$ values, we find that the resulting $Z_{\rm pred}$ are not strongly affected: RMSE~$= 0.0862$~dex.
	
These tests demonstrate that our CNN is robust to normally distributed scatter in $Z_{\rm true}$. Our results show initial promise for cases in which training data are more uncertain, such as at higher redshift. However, more testing is necessary to understand the effects of biased or correlated sources of uncertainty, and to account for the evolving relationships between metallicity and other observed properties \citeeg{Zahid2013,Salim2015}.

\subsection{Resolution and color effects} \label{sec:resolution}

\begin{figure}
	\includegraphics[width=\columnwidth]{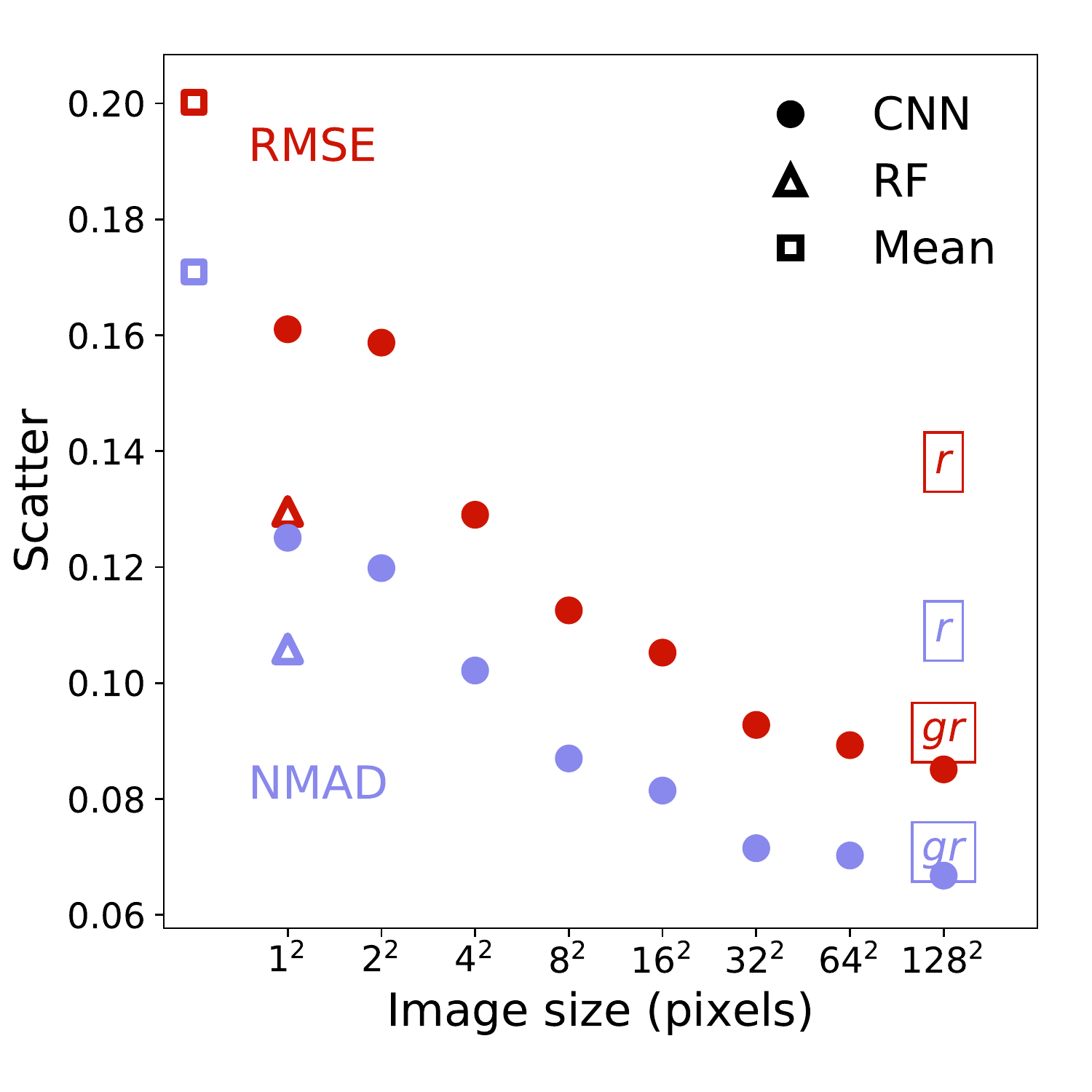}
	\caption{\label{fig:resolution}
		The effects of image resolution and color on CNN performance. Red and violet circular markers indicate scatter in the residual distribution ($\Delta Z$) measured using RMSE and NMAD, respectively, for \textit{gri} imaging. (Each point is analogous to the horizontal lines shown in Figure~\ref{fig:predicting-metallicity}.) We also show predictions from a random forest algorithm as open triangle markers, and constant $\langle Z_{\rm true}\rangle$ predictions as open square markers. Large, labeled squares indicate the scatter for images consisting of only \sdssr-band imaging and a combination of \sdssg- and \sdssr-bands.}
\end{figure}

Because our methodology is so computationally light, we can run the same CNN training and test procedure on images scaled to different sizes in order to understand the effects of image resolution. Our initial results use SDSS $38\arcsec \times 38 \arcsec$ cutouts resized to $128\times 128$ pixels, and we now downsample the same images to $64\times 64$, $32 \times 32$, $\cdots$, $2\times 2$, and even $1\times 1$ pixels via re-binning. All images retain their three channels, so the smallest $1 \times 1$ image is effectively the pixels in each of the \sdssg\sdssr\sdssi\ bands averaged together with the background and possible neighboring sources.

In Figure~\ref{fig:resolution}, we show the effects of image resolution by measuring the global scatter in $\Delta Z$ using the RMSE and NMAD metrics (shown in red and violet circular markers, respectively). Also shown is the scatter in $\Delta Z$ if we always predict the mean value of $Z_{\rm true}$ over the data set (shown using a square marker). This constant prediction effectively delivers the worst possible scatter, and the \cite{Tremonti2004} systematic uncertainty in $Z_{\rm true}$ of $\sim 0.03$~dex yields the best possible scatter. We find that both RMSE and NMAD decrease with increasing resolution, as expected if morphology or color gradients are instrumental to predicting metallicity.

There appears to be little improvement in scatter going from $1 \times 1$ to $2\times 2$ pixel images. $1\times 1$ three-color images contain similar information to three photometric data points (although because background and neighboring pixels are averaged in, they are less information-dense than photometry), which can be used to perform a crude spectral energy distribution (SED) fit. Therefore it unsurprising that the $1 \times 1$ CNN predictions perform so much better than the baseline mean prediction. A $2 \times 2$ three-color image contains four times as many pixels as a $1\times 1$ image, but because the object is centered between all four pixels, information is still averaged among all available pixels. Therefore, the scatter does not improve appreciably going from $1 \times 1$ to $2 \times 2$ resolution.\footnote{There is extra information in the $2\times 2$ pixel images in non-circularly symmetric cases. For an inclined disk, it is possible to roughly determine the orientation in the sky plane, but this information is not very useful. In the case of a major merger or interacting companion, the $2\times 2$ images may be more powerful than $1 \times 1$ images.}

The scatter is a strong function of resolution as the images are resolved from $2 \times 2$ to about $32 \times 32$ pixels. With further increasing resolution, improvement is still evident, although the scaling with scatter is noticeably weaker. Because the angular size of each image cutout stays the same, the pixel scale changes from $1\farcs184$ \perpixel\ for $32 \times 32$ images, to $0\farcs592$ \perpixel\ for $64 \times 64$ images, to $0\farcs296$ \perpixel\ for $128 \times 128$ images. The native SDSS pixel resolution is $0\farcs396$ \perpixel, such that the $64 \times 64$ and $128 \times 128$ resolutions result in the oversampling of each image. Thus, scatter is expected to plateau for images larger than $128 \times 128$. It is worth noting, however, that the CNN attempts to learn filters that depend on the size of the input image, so smaller images may result in the CNN training filters that are too low in resolution to be completely effective for prediction. Therefore, it is also not surprising that the CNN makes incremental gains for images with increasing resolution beyond $64 \times 64$ pixels.

We also train the CNN to predict metallicity using only the central $16\times16$-pixel regions of each SDSS $gri$ image. We find that the network is able to predict metallicity to within RMSE~$=0.0965$~dex. Because this value is higher than the scatter found when using the full-sized images, we conclude that the CNN loses valuable information when only the central regions are considered, and that relevant information for predicting global metallicity can be found in the galaxies' outer regions that are not probed by the 3'' SDSS spectroscopic fibers.

As a way of testing how the CNN responds to reduced color information, we have also repeated our training and testing routines using \sdssr-band and \sdssg\sdssr-band $128\times128$ images. In order to make use of our pretrained network, we  modify the original, three color JPG images to correspond to either one- or two-band SDSS images. For the single-band imaging, we duplicate the \sdssr-band data into the blue and red JPG channels. For the \sdssg\sdssr-band images, the blue and red channels correspond to the \sdssg\ and \sdssr\ filters, while the green channel is the mean of the two.

The large squares labeled ``\textit{r}'' in Figure~\ref{fig:resolution} show that the network trained and tested on single-band images performs relatively poorly (RMSE~$= 0.1381$~dex) compared to $gri$ imaging even at low resolution. The addition of a second color improves CNN performance significantly. When a second band is added to the training images (box labeled ``\textit{gr}'' in Figure~\ref{fig:resolution}), the RMSE improves to a level similar ($0.0915$~dex) to that of the original three-color images ($0.0851$~dex). This enhancement may be due to extra information that the bluer \sdssg-band provides about younger stellar populations. In both examples, the CNN is able to utilize spatial information about a galaxy to improve metallicity estimates.

\subsection{Random forest predictions for metallicity}\label{sec:RF}
We also construct a random forest (RF) of decision trees in order to predict metallicity using the implementation from \texttt{scikit-learn} \citep{Pedregosa2012}. Hyperparameters are selected according to the optimal RF trained by \cite{Acquaviva2016}. We use exactly the same data labels (\ie, galaxies) to train/validate or test the RF that we have used for training and testing the CNN, so that our measurements of scatter can be directly compared. However, we have used the \sdssg\sdssr\sdssi\ three-band photometry data (given in magnitudes) to train and predict metallicity. Since each galaxy only has three pieces of photometric information, it can be compared to the $1 \times 1$ three-band ``images'' processed by our CNN.

The RF predicts metallicity with RMSE $= 0.130$~dex, which is superior to our CNN trained and tested on $1\times 1$ and $2 \times 2$ images. This result is unsurprising because the RF is supplied aperture-corrected photometry, whereas the CNN is provided $1 \times 1 $ \sdssg\sdssr\sdssi\ ``images'' whose features have been averaged with their backgrounds. $2 \times 2$ images are only marginally more informative. When the resolution is further increased to $4 \times 4$ images, then the CNN can begin to learn rough morphological features and color gradients, which is already enough to surpass the performance (measured by both RMSE and NMAD) of the RF.
This result suggests that the CNN is able to learn a nontrivial representation of gas-phase metallicity based on three-band brightness distributions, even with extremely low-quality data.

\subsection{Comparisons to previous work}\label{sec:previous work}
CNNs have been used for a wide variety of classification tasks in extragalactic astronomy, including morphological classification \citeeg{Dieleman2015, Huertas-Company2015, 2017MNRAS.464.4420S}, distinguishing between compact and extended objects \citep{Kim2017}, selecting observational samples of rare objects based on simulations \citep{Huertas-Company2018, Lanusse2017}, and visualizing high-level morphological galaxy features \citep{Dai2018}. These works seek to improve classification of objects into a discreet number of classes, \ie, visual morphologies. Our paper uses CNNs to tackle the different problem of regression, \ie, predict values from a continuous distribution.

Examples of regressing stellar properties in the astronomical ML literature \citeeg{2000A&A...357..197B, Fabbro2018} train on synthetic stellar spectra and test on real data. Their predicted measurements of stellar properties, \eg, stellar effective temperature, surface gravity, or elemental abundance, can be derived from the available training data set. Our work is novel because we predict metallicity, a spectroscopically determined galaxy property, using only three-color images. Said another way, it is not necessarily the case that $Z$ can be predicted from our training data. However, we find that galaxy shape supplements color information in a way that is useful for predicting metallicity.

A study similar to this work is that of \cite{Acquaviva2016}, who uses a variety of machine learning methods including RFs, extremely random trees (ERTs), boosted decision trees (AdaBoost), and support vector machines (SVMs) in order to estimate galaxy metallicity. The \cite{Acquaviva2016} data set consisted of a $z \sim 0.1$ sample (with $\sim 25,000$ objects) and a $z \sim 0.2$ sample (with $\sim 3,000$ objects), each of which has five-band SDSS photometry ($ugriz$) available as inputs. These samples are sparsely populated at low metallicities, and they contain smaller fractions of objects with $Z_{\rm true} < 8.5$ than our sample, but are otherwise similarly distributed in $Z_{\rm true}$ to ours. Our samples have different sizes because we require SDSS objects to have imaging available, whereas the \cite{Acquaviva2016} criteria impose stronger spectroscopic redshift constraints.

We will first compare RF results, since this technique is common to both of our analyses, and they reveal important differences in our training data. Because outliers are defined differently in both works, we will use the RMSE metric to compare scatter between the two. \cite{Acquaviva2016} obtain RMSE of 0.081 and 0.093~dex when using RFs on the five-band photometry for the $z \sim 0.1$ and $0.2$ subsamples. Using exactly the same RF approach on a larger sample, while working with only \textit{three} bands of photometric information, we find RMSE $= 0.130$~dex. Our scatter is larger than the value reported by \cite{Acquaviva2016} by a factor of $\sim 1.5$. This result may partly be explained by the fact that \cite{Acquaviva2016} $Z_{\rm true}$ distribution is narrower than for our training data set, or the fact that our data set spans a broader range in galaxy redshift; however, some of this advantage is offset by our larger sample size.
Ultimately, it appears that the extra \sdssu\ and \sdssz\ bands supply machine learning algorithms with valuable information for predicting metallicity.

Indeed, the \sdssu\ and \sdssz-bands convey information about a galaxy's SFR and stellar mass \cite[see, e.g.,][]{Hopkins2003}. For this reason, it is possible that the RF trained on five-band photometry can estimate $Z_{\rm true}$ down to the limit of the FMR, which has very small scatter ($\sim 0.05$~dex) at fixed $M_{\star}$ \textit{and} SFR. The \sdssg{}, \sdssr{}, and \sdssi{} bands are less sensitive to the SFR, but can still provide some information about the stellar mass, and so our RF and CNN results are more linked to the MZR rather than the FMR.

Regardless of these limitations, our CNN is able to estimate metallicity with $\Delta Z = 0.085$~dex, which is comparable to the scatter in residuals using the best algorithms from \cite{Acquaviva2016}. There is evidence that the morphological information provided by using images rather than photometric data is helping the CNN perform so well: (1) the RMSE scatter decreases with increasing image resolution, and (2) it identifies edge-on galaxies as lower-$Z_{\rm pred}$ and face-on galaxies as higher-$Z_{\rm pred}$ (consistent with observational bias). Gradients in color, or identification of mergers \citeeg{Ackermann2018} may also be helpful for predicting metallicity.

\section{The mass-metallicity relation} \label{sec:MZR}
The MZR describes the tight correlation between galaxy stellar mass and nebular gas-phase metallicity. Scatter in this correlation is approximately $\sigma \approx 0.10$~dex in $Z_{\rm true}$ over the stellar mass range $8.5 < \log (\mstar / \msol) < 11.5$ \citep{Tremonti2004}, where $\sigma$ is the standard deviation of the metallicity and is equivalent to the RMSE for a normal distribution. The MZR at $z=0$ can be characterized empirically using a polynomial fit:
\begin{equation}\label{eq:mzr}
Z = -1.492 + 1.847 \log (\mstar / \msol) - 0.08026 \left [\log(\mstar / \msol)\right ]^2.
\end{equation}

The physical interpretation of the MZR is that a galaxy's stellar mass strongly correlates with its chemical enrichment. Proposed explanations of this relationship's origin include metal loss through blowout \citep[see, e.g.,][]{2002ApJ...581.1019G,Tremonti2004,Brooks2007,Dave2012}, inflow of pristine gas \cite{Dalcanton2004}, or a combination of the two \citeeg{2013ApJ...772..119L}; however, see also \cite{2013A&A...554A..58S}. Although the exact physical process responsible for the low ($0.10$~dex) scatter in the MZR is not known, its link to SFR via the FMR is clear, as star formation leads to both metal enrichment of the interstellar medium and stellar mass assembly.

The FMR connects the instantaneous ($\sim 10$ Myr) SFR with the gas-phase metallicity \citep[$\sim 1$~Gyr timescales; see, e.g.,][]{2011ApJ...734...48L} and \mstar{} (\ie, the $\sim 13$~Gyr integrated SFR). Our CNN is better suited for predicting \mstar{} rather than SFR, using the \sdssg\sdssr\sdssi bands, which can only weakly probe the blue light from young, massive stars. Therefore, we expect the scatter in CNN predictions to be limited by the MZR (with scatter $\sigma \sim 0.10$~dex) rather than the FMR ($\sigma \sim 0.05$~dex). It is possible that galaxy color and morphology, in tandem with CNN-predicted stellar mass, can be used to roughly estimate the SFR, but in this paper we will focus on only the MZR.

\subsection{Predicting stellar mass}

\begin{figure*}
	\includegraphics[width=\columnwidth]{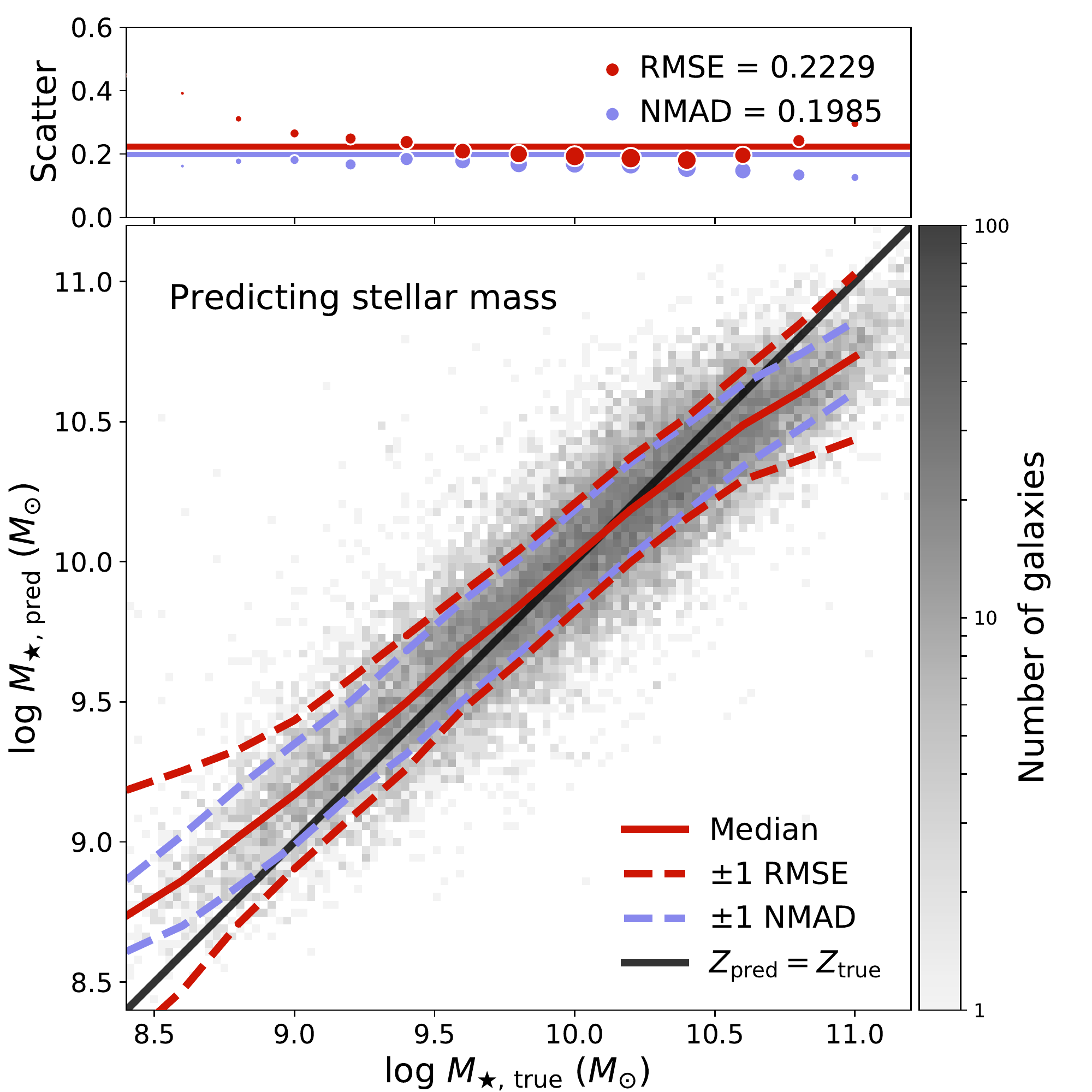}
	\includegraphics[width=\columnwidth]{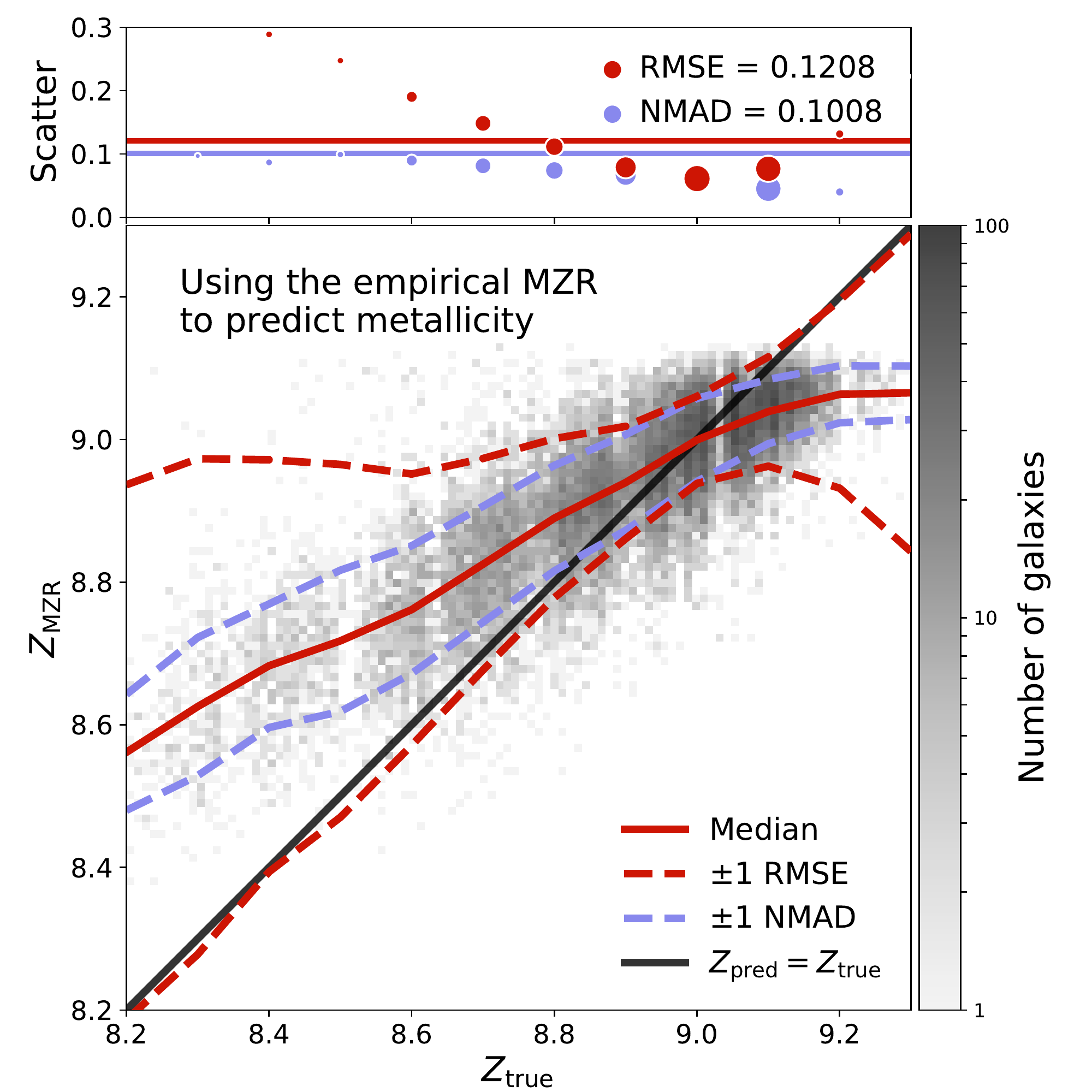}
	\caption{\label{fig:mass-metallicity}
		In the left panel, we plot the CNN predicted galaxy stellar mass against true stellar mass. Colors and marker or line styles are the same as in Figure~\ref{fig:predicting-metallicity}. In the right panel, we compare the predicted stellar mass converted to metallicity, assuming the \citet{Tremonti2004} MZR, against the true metallicity.
		These findings indicate that using the empirical MZR and CNN-predicted $M_{\star,\rm pred}$ yields poor results, unlike what we have observed in Figure~\ref{fig:predicting-metallicity}.
		}
\end{figure*}

Since galaxy stellar mass is known to strongly correlate with metallicity, and is easier to predict (than, \eg, SFR) from \sdssg\sdssr\sdssi\ imaging, we consider the possibility that the CNN is simply predicting stellar mass ($M_{\star,\rm pred}$) accurately and then learning the simple polynomial transformation in order to estimate metallicity. We can simulate this scenario by training the CNN on $M_{\star, \rm true}$ and then converting the stellar mass predictions to metallicities using Equation~\ref{eq:mzr}.

We re-run the CNN methodology to train and predict $M_{\star}$ using the 116,394 available images (out of the 142,145/142,186 original objects that have stellar mass measurements). These results are shown in the left panel of Figure~\ref{fig:mass-metallicity}. From the same subsample as before (minus three objects that do not have \mstar estimates), we verify that $M_{\rm \star, true}$ median agrees with the median of $M_{\rm \star, true}$ for values between $9.0 \lesssim \log \mstar /\msol \lesssim 10.5$. The RMSE scatter in the \mstar{} residuals is $\sim 0.22$~dex, and the NMAD is $\sim 0.20$~dex. The slope of the empirical MZR at $\log (\mstar / \msol) \sim 10$ is (0.4~dex in $Z$)/(1.0~dex in \mstar), implying that the CNN might be able to leverage the MZR and predict metallicity to $\sim 0.08$~dex (plus any intrinsic scatter in the MZR, in quadrature).

We use Equation~\ref{eq:mzr} and $M_{\star,\rm pred}$ to predict metallicity, which we call $Z_{\rm MZR}$. In the right panel of Figure~\ref{fig:mass-metallicity}, we compare $Z_{\rm MZR}$ against $Z_{\rm true}$. The scatter in residuals $Z_{\rm MZR} - Z_{\rm true}$ is $0.12$~dex, which is significantly higher than the $0.085$~dex scatter reported in Section~\ref{sec:results}. If the MZR alone were mediating the CNN's ability to estimate from
\sdssg\sdssr\sdssi\ imaging, then we would expect the scatter for $Z_{\rm pred}$ to be greater than for $Z_{\rm MZR}$; instead we find that the opposite is true. This evidence suggests that the CNN has learned to determine metallicity in a more powerful way than by simply predicting $M_{\rm \star,pred}$ and then effectively applying a polynomial conversion.

\subsection{An unexpectedly strong CNN-predicted mass-metallicity relation}
\begin{figure}
	\includegraphics[width=\columnwidth]{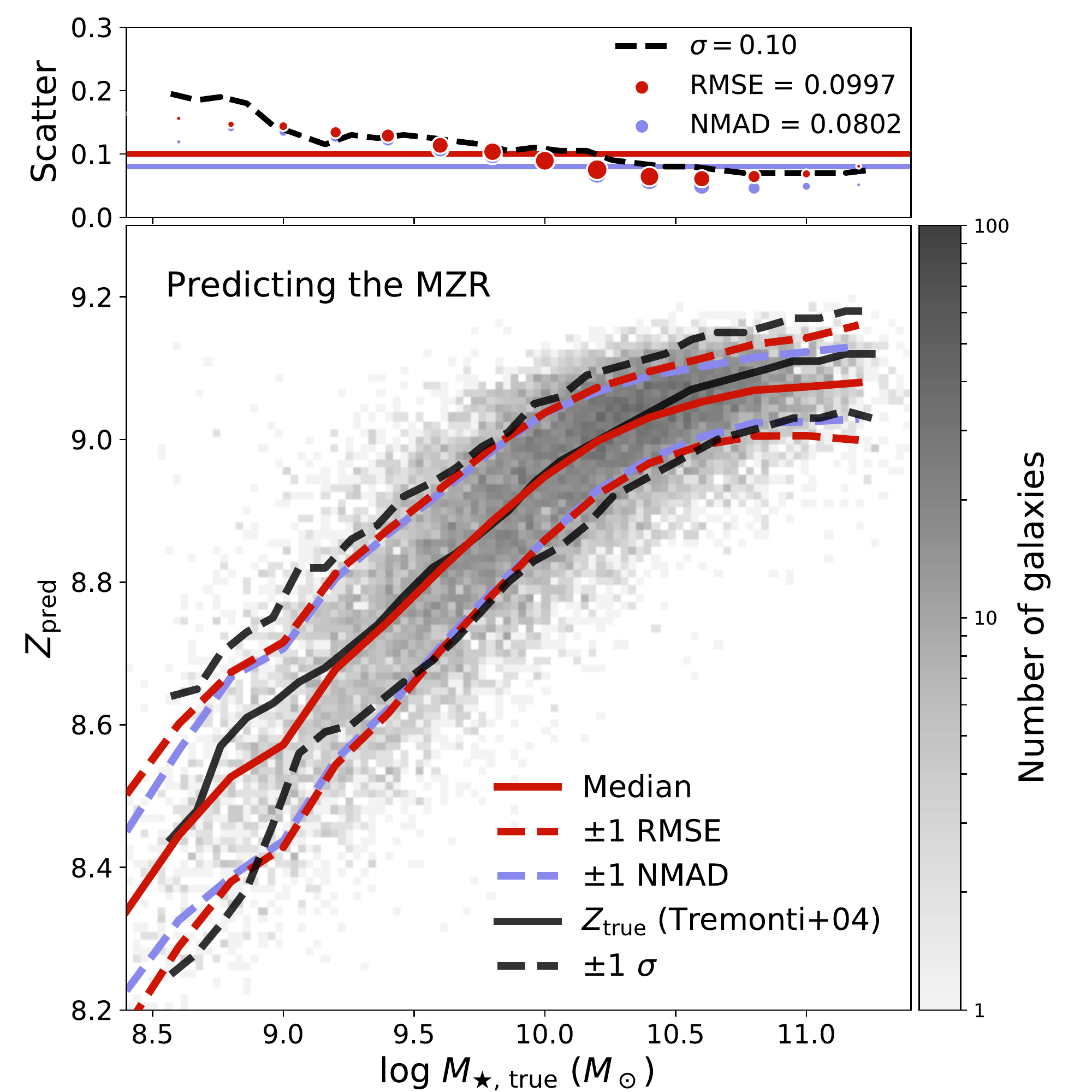}
	\caption{\label{fig:mzr}
		In the main panel, the predicted MZR comparing true \mstar\ against CNN-predicted $Z_{\rm pred}$ is shown in grayscale. The running median (solid red) and scatter (dashed red and violet) are shown in 0.2~dex mass bins. For comparison, we also show the \citet{Tremonti2004} observed median and scatter (solid and dashed black lines, respectively), which are binned by 0.1~dex in mass. In the upper panel, we show the scatter in the predicted and empirical MZR. The standard deviation of the scatter for the empirical MZR is shown as a dashed black line, while the red and violet circles respectively show RMSE and NMAD for the predicted MZR. Marker sizes are proportional to the number of galaxies in each stellar mass bin for the test data set. Global scatter in the CNN-predicted MZR appears to be comparable to, or even lower than, scatter in the empirical MZR.}
\end{figure}

The RMSE $= 0.085$~dex difference between the true and CNN-predicted metallicities can be interpreted in one of two ways: (1) the CNN is inaccurate, and $Z_{\rm pred}$ deviates randomly from $Z_{\rm true}$, or (2) the CNN is labeling $Z_{\rm pred}$ according to some other hidden variable, and $\Delta Z$ residuals represent non-random shifts in predictions based on this variable. If the first scenario were true, we would expect the random residuals to increase the scatter of other known correlations such as the MZR when predicted by the CNN. If the second were true, we would expect the scatter of such correlations to remain unchanged or shrink. We can therefore contrast the MZR constructed from $Z_{\rm pred}$ and from $Z_{\rm true}$ in order to test these interpretations.

In the main panel of Figure~\ref{fig:mzr}, we plot CNN-predicted metallicity versus true stellar mass. For comparison, we also overlay the \cite{Tremonti2004} MZR median relation and its $\pm 1~\sigma$ scatter (which is $\sim 0.10$~dex). Their empirical median relation (solid black) matches our predicted MZR median (solid red), and the lines marking observed scatter (dashed black) match $Z_{\rm pred}$ scatter as well (dashed red and violet). Over the range $9.5 \leq \log (M_{\star, \rm true}/\msol) \leq 10.5$, the RMSE scatter in $Z_{\rm pred}$ appears to be even tighter than the observed $\pm 1\,\sigma$ (dashed black). The same is true for the NMAD, which is even lower over the same interval.

In the upper panel of Figure~\ref{fig:mzr}, we present the scatter in both predicted and \cite{Tremonti2004} MZR binned by mass. We confirm that the CNN predicts a MZR that is at most equal in scatter than one constructed using the true metallicity. The stellar mass bins for which our CNN found tighter scatter than the empirical MZR are the same bins that happen to contain the greatest number of examples ($9.5 \leq \log (M_{\star, \rm true}/\msol) \leq 10.5$); thus, the strong performance of our network at those masses may be due to a wealth of training examples. If our data set were augmented to include additional low- and high-$M_{\rm \star,true}$ galaxies, then the scatter in the predicted MZR could be even lower overall.

The fact that a CNN trained on only \sdssg\sdssr\sdssi\ imaging is able to predict metallicity accurately enough to reproduce the MZR in terms of median and scatter is not trivial. The error budget is very small: $\sigma = 0.10$~dex affords only, \eg, 0.05~dex of scatter when SFR is a controlled parameter plus a 0.03~dex systematic scatter in $Z_{\rm true}$ measurements, leaving only $\sim 0.08$~dex remaining for CNN systematics, assuming that these errors are not correlated and are added in quadrature. This remaining error budget may barely be able to accommodate our result of RMSE($\Delta Z$) $= 0.085$.
Interpreting the MZR scatter as the combination of intrinsic FMR scatter, $Z_{\rm true}$ systematics, and $\Delta Z$ systematics cannot be correct since it assumes that the CNN is recovering the FMR perfectly.
As we have discussed previously, it is highly unlikely that the CNN is sensitive to the SFR, and therefore cannot probe the MZR at individual values of the SFR.

If we assume that the error budget for the MZR is not determined by the FMR, then the error ``floor'' should be $0.10$~dex.
This is immediately exceeded, as we have found RMSE $\approx 0.10$~dex for the predicted MZR without accounting for the fact that $Z_{\rm pred}$ and $Z_{\rm true}$ differ by RMSE $= 0.085$~dex!
Consider the case in which all $Z_{\rm true}$ values are shifted randomly by a Gaussian noise distribution with $\sigma = 0.085$~dex.
These shifted values should not be able to reconstruct a correlation without introducing additional scatter unless the shifts were not arbitrary to begin with.

We thus find more evidence that the CNN has learned something from the SDSS \sdssg\sdssr\sdssi\ imaging that is different from, but at least as powerful as, the MZR. One possible explanation is that the CNN is measuring some version of metallicity that is more fundamentally linked to the stellar mass, rather than $Z_{\rm pred}$ as derived from oxygen spectral lines. Another possibility is that the MZR is a projection of a correlation between stellar mass, metallicity, and a third parameter, perhaps one that is morphological in nature. If this is the case, then the \cite{Tremonti2004} MZR represents a relationship that is randomly distributed in the yet unknown third parameter, while our CNN would be able to stratify the MZR according to this parameter (much as the FMR does with the SFR). We are unfortunately not able to identify any hidden parameter using the current CNN methodology, but we plan to explore this topic in a future work.

\section{Summary}\label{sec:summary}
We have trained a deep convolutional neural network (CNN) to predict galaxy gas-phase metallicity using only $128 \times 128$-pixel, three-band (\sdssg\sdssr\sdssi), JPG images obtained from SDSS. We characterize CNN performance by measuring scatter in the residuals between predicted ($Z_{\rm pred}$) and true ($Z_{\rm true}$) metallicities.
Our conclusions are as follows:

\begin{enumerate}
	\item By training for a half-hour on a GPU, the CNN can predict metallicity well enough to achieve residuals characterized by RMSE $= 0.085$~dex (or outlier-insensitive NMAD $= 0.067$~dex). These findings may be promising for future large spectroscopy-limited surveys such as LSST.

	\item We find that the residual scatter decreases in an expected way as resolution or number of channels is increased, suggesting that the CNN is leveraging both the spatial information about a galaxy's light distribution and the color in order to predict metallicity.

	\item The CNN outperforms a random forest trained on $gri$ photometry if provided images larger than $4\times 4$ pixels, and is as accurate as a random forest trained on $ugriz$ photometry when given $128 \times 128$ pixel \sdssg\sdssr\sdssi\ images.

	\item We find that scatter in the mass-metallicity relation (MZR) constructed using CNN-predicted metallicities is as tight as the empirical MZR ($\sigma = 0.10$~dex).	Because predicted metallicities differ from the ``true'' metallicities by RMSE $= 0.085$~dex, the only way that the predicted MZR can have such low scatter is if the CNN has learned a connection to metallicity that is more strongly linked to the galaxies' light distributions than their nebular line emission.
\end{enumerate}

All of the code used in our analysis and for making the figures can be accessed at \url{https://github.com/jwuphysics/galaxy-cnns}. 

\section*{Acknowledgements}

SB is supported by NSF Astronomy and Astrophysics Research Program award number 1615657.
The authors thank the anonymous referee for useful comments that have improved this paper, particularly in terms of its scientific content.
The authors thank Andrew Baker, Eric Gawiser, and John Hughes for helpful comments and discussions, and also thank David Shih and Matthew Buckley for the use of their GPU cluster in the Rutgers University Experimental High Energy Physics department. 
JW thanks Jeremy Howard, Rachel Thomas, and the development team for creating the \texttt{fastai} on-line courses and deep learning library.\footnote{\url{https://github.com/fastai/fastai}}
JW also thanks Florian Peter for valuable assistance with using the \texttt{fastai} library.
This research made use of the {\tt IPython} package \citep{Perez2007} and {\tt matplotlib}, a Python library for publication quality graphics \citep{Hunter2007}.

Funding for the Sloan Digital Sky Survey IV has been provided by the Alfred P. Sloan Foundation, the U.S. Department of Energy Office of Science, and the Participating Institutions. SDSS-IV acknowledges
support and resources from the Center for High-Performance Computing at
the University of Utah. The SDSS web site is \url{www.sdss.org}.

SDSS-IV is managed by the Astrophysical Research Consortium for the
Participating Institutions of the SDSS Collaboration including the
Brazilian Participation Group, the Carnegie Institution for Science,
Carnegie Mellon University, the Chilean Participation Group, the French Participation Group, Harvard-Smithsonian Center for Astrophysics,
Instituto de Astrof\'isica de Canarias, The Johns Hopkins University,
Kavli Institute for the Physics and Mathematics of the Universe (IPMU) /
University of Tokyo, the Korean Participation Group, Lawrence Berkeley National Laboratory,
Leibniz Institut f\"ur Astrophysik Potsdam (AIP),
Max-Planck-Institut f\"ur Astronomie (MPIA Heidelberg),
Max-Planck-Institut f\"ur Astrophysik (MPA Garching),
Max-Planck-Institut f\"ur Extraterrestrische Physik (MPE),
National Astronomical Observatories of China, New Mexico State University,
New York University, University of Notre Dame,
Observat\'ario Nacional / MCTI, The Ohio State University,
Pennsylvania State University, Shanghai Astronomical Observatory,
United Kingdom Participation Group,
Universidad Nacional Aut\'onoma de M\'exico, University of Arizona,
University of Colorado Boulder, University of Oxford, University of Portsmouth,
University of Utah, University of Virginia, University of Washington, University of Wisconsin,
Vanderbilt University, and Yale University.

\bibliographystyle{mnras}

\appendix
\section{Convolution neural network details}

\subsection{Residual neural network architecture}
CNNs are divided into ``layers'' that compute the convolutions of filters, or kernels, with each of the inputs.
A Rectified Linear Unit (ReLU) activation function is applied to the convolved output \citep[ReLUs have been shown to propagate information about the relative importances of different features, and are effective for training deep neural networks;][]{Nair2010}.
In a residual CNN, multiple convolutional layers containing small (e.g., $3\times 3$) filters are arranged sequentially, and a final ``shortcut connection'' adds the first layer, unaltered, to the final output \citep[before the final ReLU activation; see, e.g., Figure~2 of][]{He2015}.
Such combinations of convolutions, activations, and shortcuts are called residual building blocks.

We use a 34-layer residual convolutional neural network with the architecture described by \cite{He2015}, and implemented using \texttt{PyTorch} \citep[version 0.3.1;][]{pytorch} provided by the \texttt{fastai} framework \citep[version 0.7;][]{fastai}.
A full description of the architecture's layers can be found in the online \texttt{PyTorch} documentation,\footnote{\url{https://pytorch.org/docs/stable/torchvision/models.html}} but we also provide a brief overview below.

The resnet can be separated into three ``layer groups'' that roughly correspond to the levels of abstraction able to be learned by the network. Once an image is fed into the network, \eg, a three-channel $128\times 128$ SDSS image, it is effectively converted into activation maps that depend on how well the filters match the input image. These maps are further convolved with the next layer of filters, and this process continues until the last layer group is reached. The activation maps are periodically downsampled, max pooled, or average pooled, which effectively halve the map sizes in each spatial dimension \citep[for more about pooling layers in CNNs, see][]{Scherer2010}. The first two layer groups comprise multiple residual building blocks, and the final layer group consists of two fully connected linear layers, with a ReLU activation after the first and no activation after the second. The last fully connected layer does not have an activation function because we are working on a regression problem, and so the weights trained in that layer should be tuned to predict metallicity in the desired range.

\subsection{Adaptive learning rates}
Neural network performance tends to depend dramatically on choice of hyperparameters. After an image is fed forward and the residual (= prediction $-$ true) value is computed, relative contributions of error are propagated backward through the network, starting from the final layer and ending at the first layer. Using gradient descent of the loss (in our case, the root mean squared error), the network layers' weights are adjusted according to their error contributions multiplied by the \textit{learning rate}. The process of computing errors from known images and metallicities and updating weights is called \textit{training}, and when all of the training data set has been used to adjust network weights, a training \textit{epoch} is completed.

The learning rate can be thought of as the step size during each weight update. A high learning rate allows the network to improve quickly, but at some point the large step size may become too coarse for additional optimization; conversely, a low learning rate might allow the network to traverse every bump and wiggle in the error landscape, but might also take a very long time to reach convergence (or get stuck indefinitely in a local minimum). We first select a learning rate by using the method described by \cite{CLR}. Over a number of epochs, the learning rate is reduced (or \textit{annealed}) as the network needs to make more fine-tuned updates in order to achieve better accuracy. We use a method called cosine annealing, during which the learning rate is annealed with the cosine function continuously over individual (or batches of) training examples.
It has been shown that if the learning rate is annealed and then restarted after one or more epochs, the network is less likely to get caught in local minima and overall accuracy is improved. We refer to \cite{SGDR} for details about employing cyclical learning rates and gradient descent with restarts, which are implemented in our CNN.

\subsection{Optimization techniques and preventing overfitting} \label{sec:optimization}

Losses are computed for small ``batches'' of training examples at a time. Gradients that minimize each batch are expected to be noisier than gradients that are computed to optimize the entire training data set loss. This technique of \textit{stochastic gradient descent} helps prevent the CNN from overfitting training data, which is a possibility given the huge number of parameters in a deep CNN. We also use weight decay, another commonly used regularization technique, which adds a decay term proportional to each layer weight during the update step of training \citeeg{Krogh1992}.

As the learning rate is annealed with increasing numbers of batches, the weight updates are also expected to diminish.
The Adam optimizer adaptively smooths the gradient descent in a way that depends on previous gradients \citep{Kingma2014}.
Adam is analogous to rolling downhill with gravitational potential, momentum, and friction terms (whereas gradient descent would be analogous to movement dependent only on the potential at its given time step).
For caveats about combining weight decay and Adam, see \cite{Loshchilov2017}, whose updated algorithm is implemented in \texttt{fastai}.

We implement batch normalization (BN), a technique developed to fix a problem that previously caused deep networks to train extremely slowly \citep{batchnorm}. To briefly recap the issue: updates to the layer weights depend on the contribution of the backpropagated error, but when the number of layers is large (\ie, in a deep CNN), the contribution becomes vanishingly small. BN is simply the rescaling of each input to the nonlinear activation so that it has mean of zero and standard deviation of unity (\ie, subtract the mean and divide by the standard deviation). A new choice of hyperparameter is the batch size, or the number of training examples from which the mean and variance are calculated); we choose 256 based on tests of performance in ten training epochs.

Dropout is a method of disabling a random subset of connections after linear layers in a network in order to improve the network's generalizability \citep{dropout}. The ensemble of learned gradients is less prone to overfit the training data set because the network is forced to discard random (and potentially valuable) information. The resulting network is better able to, \eg, learn subtle differences in the data that would otherwise be ignored when more obvious features dominate the gradient descent process. We apply dropout layers only to the final fully connected layers in our deep CNN, and avoid dropout in the batch-normalized layers \citep[as recommended by][]{batchnorm}. We use dropout rates of 0.25 for the linear layer after the early group, and 0.50 at the later linear layer, both of which are \texttt{fastai} defaults.

\subsection{Training the network}
We initialize the network using weights that have been pretrained on the 1.7~million example ImageNet data set \citep[which contains 1000 classes of objects;][]{ImageNet}. The network should more quickly optimize toward the global minimum loss through transfer of low-level features already learned in earlier layers of the network \citep[known as transfer learning; see, \eg,][]{Pan2010}.

We train only the final layer group for the first two epochs, which can be accomplished by not updating weights in the first two layer groups. The learning rate is initially set to 0.1 and then annealed according to a cosine schedule over an epoch (and then restarted to 0.1 at the beginning of the following epoch). We then allow the updating of weights in all layer groups while setting the learning rates to 0.01, 0.03, and 0.1 for the first, second, and last layer groups, respectively. This approach allows the final group of fully connected layers to respond strongly to different types of training examples (\eg, galaxies that appear very different in \sdssg\sdssr\sdssi\ imaging) while the earlier layers are trained very slowly in order to preserve their more general features. Using these layered learning rates, we train the full network using a cosine annealing schedule that spans one, one, two, and then four epochs (where the different learning rates are annealed by the same amount). Using this combination of learning rate schedules, we find that our network quickly achieves low training losses (RMSE $\sim 0.085$ on validation data sets). Altogether, only ten epochs of training are needed, which takes under 30 minutes on a GPU. We find that further training does yield some gains, but this improvement plateaus around RMSE $\sim 0.083$ and takes many more hours.

\subsection{Data augmentation}\label{sec:data aug}
Nearly all neural networks benefit from larger training samples because they help prevent overfitting. Beyond the local Universe, galaxies are seen at nearly random orientation; such invariance permits synthetic data to be generated from rotations and flips of the training images \citep[see, \eg,][]{2014arXiv1409.1556S}. Each image is fed into the network along with four augmented versions, thus increasing the total training sample by a factor of five.

This technique is called data augmentation, and is particularly helpful for the network to learn uncommon truth values (\eg, in our case, very metal-poor or metal-rich galaxies). Each augmented image is fed-forward through the network and gradient contributions are computed together as part of the same batch. A similar process is applied to the network during predictions, which is known as test-time augmentation (TTA), whereby synthetic images are generated according to the same rules applied to the training data set. The CNN predicts an ensemble average over the augmented images, which tends to further improve RMSE by a few percent. We use the default hyperparameters in the \texttt{fastai} library.

\bsp	
\label{lastpage}
\end{document}